\documentclass[twocolumn,floatfix,showpacs,prl,aps,tightenlines
]{revtex4}
\usepackage{graphicx}
\usepackage{color}
\usepackage{psfrag}
\usepackage{dcolumn}
\usepackage{bm}

\newcommand{\bea}{\begin{eqnarray}}
\newcommand{\eea}{\end{eqnarray}}
\newcommand{\beq}{\begin{equation}}
\newcommand{\eeq}{\end{equation}}
\newcommand{\KMS}{\rm km\,s^{-1}}

\newcommand{\mysection}[1]{\noindent {\it #1}: }
\begin{document}

\title{Hangup Kicks: Still Larger Recoils by Partial Spin/Orbit
Alignment of Black-Hole Binaries}

\author{Carlos O. Lousto}
\affiliation{Center for Computational Relativity and Gravitation,
School of Mathematical Sciences,
Rochester Institute of Technology, 85 Lomb Memorial Drive, Rochester,
 New York 14623, USA}

\author{Yosef Zlochower} 
\affiliation{Center for Computational Relativity and Gravitation,
School of Mathematical Sciences,
Rochester Institute of Technology, 85 Lomb Memorial Drive, Rochester,
 New York 14623, USA}

\date{\today}

\begin{abstract} 
We revisit the scenario of the gravitational radiation recoil acquired
by the final remnant of a black-hole-binary merger by studying a set
of configurations that have components of the spin both aligned with
the orbital angular momentum and in the orbital plane. We perform a
series of 42 new full numerical simulations for equal-mass and
equal-spin-magnitude  binaries.  We extend previous recoil fitting
formulas to include nonlinear terms in the spins and successfully
include both the new and known results. The  new predicted maximum
velocity approaches 5000km/s for spins partially aligned with the
orbital angular momentum, which leads to an important increase of the
probabilities of large recoils in generic astrophysical mergers. We
find non-negligible  probabilities for recoils of several thousand
km/s from accretion-aligned  binaries.
\end{abstract}

 \maketitle

\mysection{Introduction}
With the breakthroughs of 2005 in the numerical techniques to evolve
black-hole binaries (BHBs) \cite{Pretorius:2005gq, Campanelli:2005dd,
Baker:2005vv}, Numerical Relativity (NR) became a very important
tool to explore highly-dynamical and nonlinear predictions of
General Relativity. In the last few years we have gained notable
insight into the modeling of gravitational radiation to assist
laser interferometer detectors \cite{Aylott:2009ya}. There are also numerous
examples of explorations in the realm of Mathematical Relativity
and in the astrophysical scenarios for supermassive
black-hole mergers and retention of black holes in galaxies and
globular clusters \cite{Campanelli:2007ew,Volonteri:2010hk}.

Some of the most striking recent discoveries are related to effects
due to the intrinsic spin of the individual BHs during the final merger 
stage of BHBs.  In general,
the spins and orbital plane precess during the inspiral and the spin
direction of the remnant is misaligned with the individual BH
spins~\cite{Campanelli:2006fy} prior to merger (spin flips). Spins can
also have a dramatic effect on the inspiral rate. When the BH spins
are partially aligned with the orbital angular momentum, the merger is
delayed, while when they are antialigned, the merger happens much more
quickly. This ``hangup'' effect~\cite{Campanelli:2006uy} is due to an
unexpectedly strong spin-orbit coupling.  Perhaps even more
surprisingly are
the very large recoils~\cite{Campanelli:2007ew} acquired by the remnant
of the merger of comparable-masses, highly-spinning, BH in the
``superkick'' configuration, where the spins lie along the orbital plane
(equal magnitude, but opposite in direction)
\cite{Campanelli:2007ew,Campanelli:2007cga,Gonzalez:2007hi}.  Initial
studies, which indicated that these BHBs recoil  at up to 
$\sim 4000\ \KMS$~\cite{Campanelli:2007cga,Lousto:2010xk}, prompted
astronomers to search for possible recoil candidates.  To date,  a few
interesting cases of galaxies with cores  displaying differential
radial velocities of several thousands km/s \cite{Komossa:2008qd,
Civano:2010es,Shields:2009jf} have been found. More systematic recent
studies produced tens of potential candidates
\cite{Tsalmantza:2011ju,Eracleous:2011ua}.

The discovery of large recoils also triggered theoretical statistical studies of
BHB dry mergers \cite{Lousto:2009ka} and Monte Carlo simulations of wet
premergers to study the effect of accretion and resonances on spin
distributions~\cite{Schnittman:2004vq, Kesden:2010yp,Kesden:2010ji}.
Accretion tends to align spins with the orbital angular momentum
\cite{Bogdanovic:2007hp,Dotti:2009vz}; resulting in a notable reduction in the
probabilities of observing large recoils, either directly, or through their 
influence on the galactic cores, for small and medium
sized galaxies \cite{Volonteri:2010hk}. For BHs with masses
larger than $10^8 M_\odot$, alignment by accretion is less effective.
In this Letter we revisit the scenarios for the generation of
recoils by studying a set of configurations that combine two of the 
largest spin effects observed in BHBs, the hangup effect and superkicks. 
The combined effect appears to be  a dramatic increase in the probability
distribution for large recoils, as we will show below. 

\mysection{Full Numerical Simulations}
We evolved a set of 42 equal-mass, spinning, quasicircular configurations using
the {\sc
LazEv}~\cite{Zlochower:2005bj} implementation of the moving puncture
formalism~\cite{Campanelli:2005dd,Baker:2005vv}, with the conformal
factor $W=\sqrt{\chi}=\exp(-2\phi)$ suggested by~\cite{Marronetti:2007wz}
as a dynamical variable.
For the runs presented here
we use centered, eighth-order finite differencing in
space~\cite{Lousto:2007rj} and a fourth-order Runge-Kutta time integrator. 
The {\sc
LazEv} code used the {\sc Cactus}/{\sc EinsteinToolkit}~\cite{cactus_web,
einsteintoolkit} numerical infrastructure along with the
{\sc Carpet}~\cite{Schnetter-etal-03b} mesh refinement driver.
We use the
{\sc
TwoPunctures}~\cite{Ansorg:2004ds} thorn to calculate the initial data.
We use {\sc AHFinderDirect}~\cite{Thornburg2003:AH-finding} to locate
apparent horizons.  We measure the magnitude of the horizon spin using
the Isolated Horizon algorithm detailed in~\cite{Dreyer02a}.

\begin{figure}
  \includegraphics[width=2.0in,height=1in]{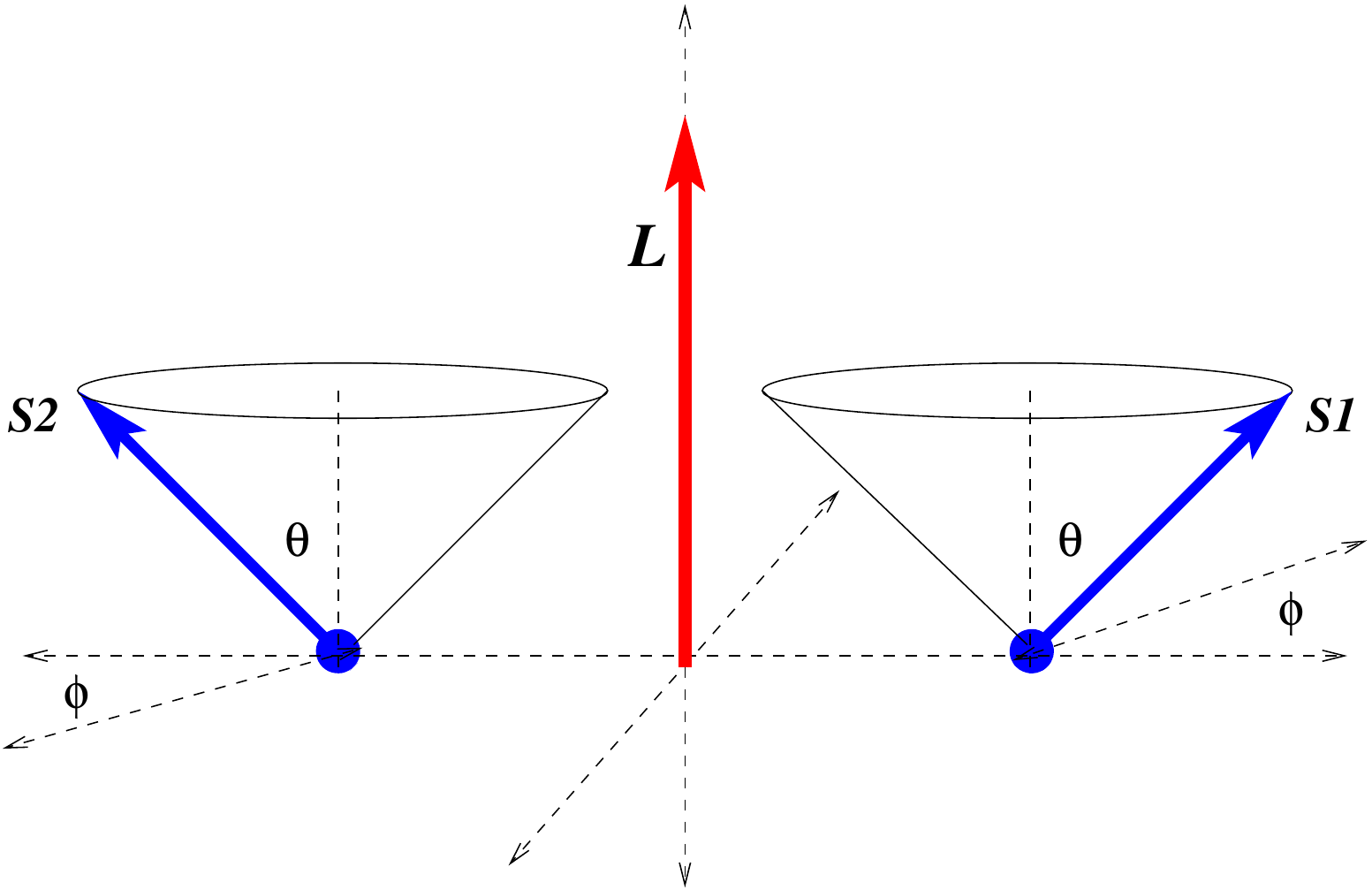}
  \caption{Black-hole-binary configuration for hangup recoils.} 
\label{fig:config}
\end{figure}

Our configurations have the property that the in-plane components of
the spins of the two BHs have the same magnitude, but opposite signs, 
while the out-of-plane
components have the same magnitude and sign (see
Fig.~\ref{fig:config}). They thus combine the
hang-up~\cite{Campanelli:2006uy} and 
superkick~\cite{Campanelli:2007ew, Campanelli:2007cga} effects.
Additionally, the orbital plane does not precess, but rather moves up
and down along the direction of orbital angular momentum as the binary evolves.

We performed a set of 30 simulations with individual BH spins
of magnitude $\alpha=1/\sqrt{2}$ and $12$ simulations with BH spin magnitudes
 of $\alpha=0.91$, where $\vec \alpha$ is the
normalized
spin of the BH.
The $\alpha=1/\sqrt{2}$ configurations were split into five sets of 6,
where the
runs in each individual set had the same initial angle $\theta$ 
between
the spin direction and orbital angular momentum direction (here we
chose $\theta=22.5^\circ$, $45^\circ$, $60^\circ$, $120^\circ$, $135^\circ$). In each
set with a given $\theta$, we chose the initial orientation
 $\phi_i$ between the in-plane
spin and linear momentum to be $0^\circ$, $30^\circ$, $90^\circ$,
$130^\circ$, $210^\circ$, and $315^\circ$. For the $\alpha=0.91$ runs,
we used the same initial 6 $\phi_i$ configurations for $\theta=60^\circ$ 
and $\theta=15^\circ$.
 We combine
these results with the simulations of~\cite{Lousto:2010xk} (which have
$\theta=90^\circ$) in order
to perform our analysis below.

We set up the initial separations, such that each binary completed
5-6 orbits, prior to merger (to reduce eccentricity). The initial
separations varied between 10.16M and 8.2M (depending on the magnitude
of the hangup effect). 

\mysection{Results and Analysis}
In a previous study~\cite{Lousto:2010xk},
 we found that the superkick recoil (where the two BHs have equal
mass, equal intrinsic spin magnitudes $\alpha$, and spins lying in the orbital plane
in opposite directions)  has the
following dependence on spin $\alpha$ and orientation $\phi$ (the
angle between the in-plane spin vector and the infall direction
near merger),
\begin{eqnarray}
  V &=& V_1 \cos(\phi-\phi_1) + V_3 \cos(3\phi - 3\phi_3), \nonumber \\
    V_1 &=& V_{1,1} \alpha + V_{1,3}\alpha^3, \nonumber \\
    V_3 &=& V_{3,1} \alpha + V_{3,3} \alpha^3,
\label{eq:superkick}
\end{eqnarray}
where
$V_{1,3}=(-15.46\pm2.66)\ \KMS$, $V_{3,1}=(15.65\pm3.01)\ \KMS$, and
$V_{3,3}=(105.90\pm4.50)\ \KMS$, while $V_{1,1} = (3681.77\pm2.66)\ \KMS$.
From that study, it was clear that in the superkick configuration,
the dominant contribution, even at large $\alpha$, is linear
in $\alpha$ and proportional to $\cos(\phi)$. Note that because
of the small contributions of $V_3$ and $V_{1,3}$, we neglect these
terms in the statistical studies below (where we take a uniform
distribution in $\phi-\phi_1$).
 
\begin{figure}
  \caption{A fit of the recoil ($V_1$) to the form
Eq.~(\ref{eq:pade}) for the $\alpha=1/\sqrt{2}$ configurations,
and predictions (based on this fitting) for the
$\alpha=0.91$ recoils. Note how well the $\alpha=0.91$ curve matches
the three measured values.
For reference, curves corresponding to the original empirical 
formula prediction (which only had terms linear
in $\Delta$) for $\alpha=1/\sqrt{2}$ and the new
formula for $\alpha=1$ are also included. Note the skew in the
velocity profile compared to the linear predictions.
}
  \includegraphics[width=3in,height=2in]{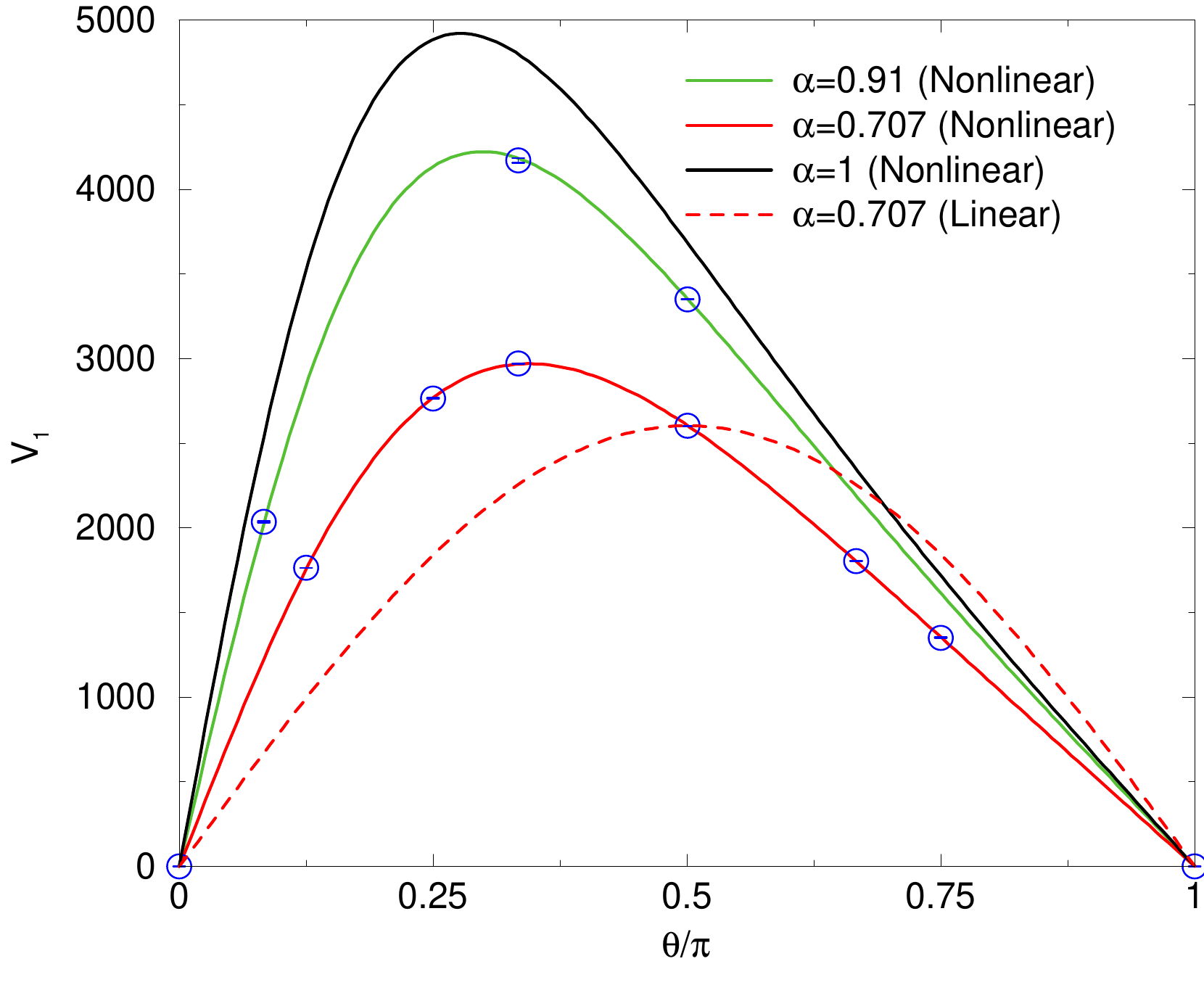}
\label{fig:fit}
\end{figure}

Our initial motivation for the current study was to determine if the hangup
effect~\cite{Campanelli:2006uy}, which amplifies the amount of
radiation emitted by the BHB, also affects the maximum recoil. To this
end, we looked at configurations  that combined both
the hangup and superkick effects.
  Based on the superkick formula~(\ref{eq:superkick}),
 we expected that the
recoil would have the form
\begin{eqnarray}
  V_1 = V_{1,1} \alpha \sin \theta + A \alpha^2 \sin\theta \cos \theta
+ \nonumber\\
      B \alpha^3 \sin\theta \cos^2\theta  + C \alpha^4 \sin\theta
\cos^3\theta,
  \label{eq:FS}
\end{eqnarray} 
where $V_1$ is the component of the recoil proportional to $\cos
\phi$,  $V_{1,1}$ arises from the superkick formula, and the
remaining terms are proportional to linear, quadratic, and higher
orders in $S_z/m^2=\alpha \cos \theta$ (the spin component in the
 direction of the orbital angular momentum). Here, we do not consider terms
higher-order in the in-plane component of $\vec \Delta\propto \vec \alpha_2
- q \vec \alpha_1$ denoted by $\Delta^\perp$
($\Delta^\perp\propto\alpha \sin \theta$ here), where $q=m_1/m_2$ is the mass
ratio,
because our previous studies
showed that these terms were small at $\theta=90^\circ$. A fit to this
ansatz~(\ref{eq:FS}) showed that
the coefficients $V_{1,1}=(3677.76\pm15.17)\ \KMS$,
$A=(2481.21\pm67.09)\ \KMS$,
$B=(1792.45\pm92.98)\ \KMS$,
$C=(1506.52\pm286.61)\ \KMS$
 converge very slowly and have relatively large uncertainties. 
In addition, we propose
the modification
\begin{equation} 
  V_1 = D \alpha \sin \theta\, \left(\frac{1 + E \alpha \cos
\theta}{1+F \alpha \cos \theta}\right),
\label{eq:pade}
\end{equation} 
which can be thought of as a resummation of Eq.~(\ref{eq:FS}) with an
additional term $E \alpha \cos \theta$,
and fit to $D$, $E$, $F$ (where we used the prediction
of~\cite{Lousto:2010xk} to model the  $V_1$ for $\theta=90^\circ$)
and find $D=(3684.73\pm5.67)\ \KMS$, $E=0.0705\pm0.0127$, and
$F=-0.6238\pm0.0098$. Note that $E$ is approximately $1/10$ of $F$,
indicating that corrections to this formula converge quickly. The two
formulas (\ref{eq:FS}) and (\ref{eq:pade}) give very similar results
for a broad range of $\alpha$.
 We then use Eq.~(\ref{eq:pade}) to predict the
recoil for higher spin $\alpha=0.91$ and test this formula for three angles
$\theta=90^\circ$, $\theta=60^\circ$, and $\theta=15^\circ$, with very good
agreement (see Fig.~\ref{fig:fit}). In actuality,
both Eq.~(\ref{eq:pade}) and Eq.~(\ref{eq:FS}) provide
accurate predictions for our measured recoils at $\alpha=0.91$.
 The results are startling. The
recoil is not maximized at $\theta=90^\circ$, as was previously
assumed based on linear spin-orbit PN expressions, but rather at smaller angles (see Table~\ref{tab:max_ang}).
Additionally, the maximum recoil is closer to $5000 \KMS$.
Note that we measured recoils from actual simulations as large as $4171 \KMS$ for
$\alpha=0.91$, 
which is larger than the previously predicted maximum possible recoil
of $3681 \KMS$ for quasicircular binaries. This can have profound
astrophysical implications because partial alignment of the binary
(e.g.\ due to accretion),
rather than inhibiting large recoils, can actually amplify them, leading to
much larger probabilities for observing high recoils.

\begin{table}
\caption {The angle $\theta$ that gives the largest recoil along
with the magnitude of this recoil for different values of
the individual BH spin $\alpha$. The columns to the left
are from Eq.~(\ref{eq:pade}), while the columns to the right
are for Eq.~(\ref{eq:FS})}
\label{tab:max_ang}
\begin{ruledtabular}
\begin{tabular}{lcc|cc}
$\alpha$ & $\theta_{\rm max}$ & $V_{\rm max}$ & $\theta_{\rm max}$ &
$V_{\rm max}$\\
\hline
0.1    & $86.02^\circ$ & 369.36 & $86.13^\circ$ & 368.62 \\
0.5    & $70.16^\circ$ & 1961.38 & $69.99^\circ$ & 1955.51\\
$1/\sqrt{2}$ & $61.90^\circ$ & $2968.71$ & $61.33^\circ$ & 2967.85\\
0.91   &$53.55^\circ$ & $4224.93$ & $53.92^\circ$ & 4231.93\\
1   & $49.67^\circ$ & $4925.94$ & $51.22^\circ$ &4915.22\\
\end{tabular}
\end{ruledtabular}
\end{table}

Using the same post-Newtonian analysis~\cite{Racine:2008kj} as in~\cite{Lousto:2009mf}, 
we can extend 
formulas~(\ref{eq:FS})~and~(\ref{eq:pade}) to less
symmetric configurations by replacing $\alpha \sin\theta$ by
$[\alpha_2^\perp - q \alpha_1^\perp]/(1+q)$ and $\alpha \cos \theta$
by $2[\alpha_2^z + q^2 \alpha_1^z]/(1+q)^2$.
Importantly, we are assuming that terms
proportional to $[\alpha_2^\perp - q \alpha_1^\perp]^n$ (for $n>1$)
are negligible. This can be verified by confirming that 
formulas~(\ref{eq:FS})~and~(\ref{eq:pade}) are accurate for
all $\theta$ and  $\alpha$ (a subject of our ongoing analysis that
will be reported in a forthcoming paper).
We emphasize that the proposed extension is an ansatz, that while
reasonable as a starting point for the modeling, needs to be
thoroughly tested and refined. 
In this letter we will use  Eq.~(\ref{eq:FS}) when generalizing
to unequal masses and arbitrary spin orientations.
A generalization of the resummation~(\ref{eq:pade}) to the generic
mass ratio $q$ case is nontrivial, and will be discussed in an upcoming paper.
Our ansatz for the generic recoil (for brevity, we only display
the term in the direction of the orbital angular momentum 
proportional to $\Delta_\perp$) then is
\begin{eqnarray}
  v_\|&=& 16 \frac{\eta^2}{(1+q)}
    \left|\alpha_2^\perp-q\alpha_1^\perp\right|
 \left[V_{1,1} + A \tilde S + B \tilde S^2  + C \tilde S^3
\right]\times\nonumber \\
&& \cos(\phi_\Delta - \phi_1),
\label{eq:generic}
\end{eqnarray}
where
$\tilde S = 2(\alpha^z_2 + q^2 \alpha^z_1)/(1+q)^2$,
$\phi_\Delta$ is angle between $\vec \Delta_\perp\propto
\vec \alpha_2^\perp - q \vec \alpha_1^\perp$ and the infall direction
(evaluated at a fiducial point around merger),
$\phi_1$ is a constant, and $\eta = q/(1+q)^2$ is the symmetric
mass ratio. See Ref.~\cite{Lousto:2009ka} for the remaining components
of the recoil velocity.

\begin{figure}
\caption{The recoil probability distribution using the new and old
empirical formulas for the recoil, starting from a distribution of
BHB configurations consistent with recent models for
accreting binaries~\cite{Volonteri:unpub}.
The new formula predicts a significantly larger probability for
high recoils. }
\includegraphics[width=3.0in,height=1.5in]{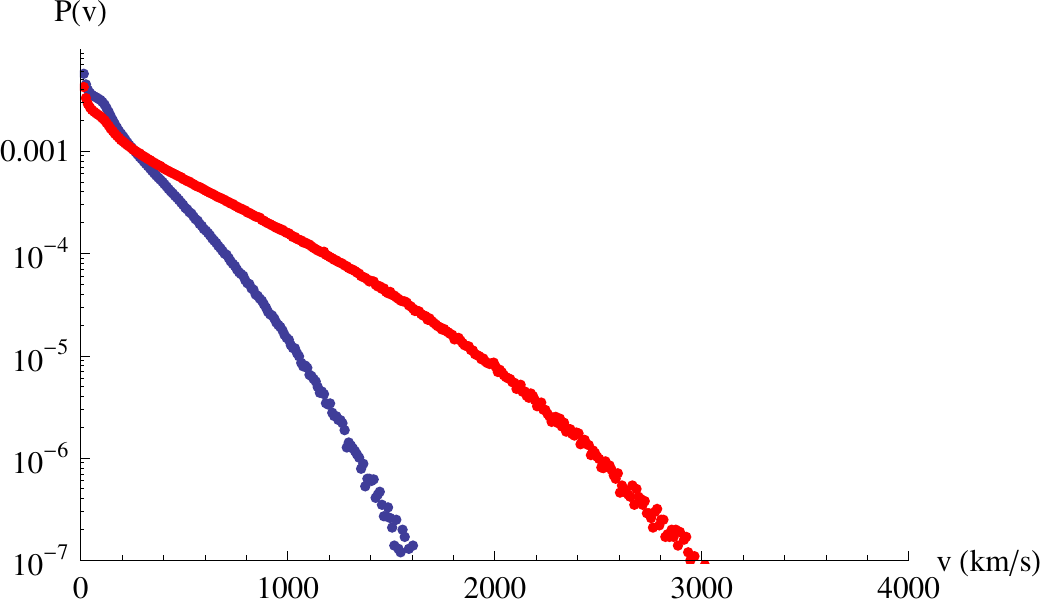}
\label{fig:rec}
\end{figure}

\mysection{Astrophysical Implications}
To test the effect of our new empirical formula on the predicted
recoil rates, we consider a model distribution of BHBs with 
mass ratio distribution
$P(q) \propto q^{-0.3} (1-q)$~\cite{Yu:2011vp, Stewart:2008ep,
Hopkins:2009yy}, spin distribution $P(\alpha)
\propto (1-\alpha)^{(b-1)} \alpha^{(a-1)}$,
with parameters $a=4.8808$ and $b=1.72879$ (which models the
distribution kindly provided by M.~Volonteri~\cite{Volonteri:unpub}
 resulting from spin-up effects due
to accretion), spin-direction distribution
$P(\theta) \propto (1-\theta)^{(b-1)} \theta^{(a-1)}$, with
parameters $a=2.5$ and $b=7$ (which approximates the spin-direction
distribution provided by M.~Volonteri~\cite{Volonteri:unpub},
 which is a simplified
model for the spin distribution for hot accretion~\cite{Dotti:2009vz}),
and $P(\phi)$ uniform in $0\leq \phi \leq 2\pi$. We model $10^7$ BHBs consistent
with these distributions and measure the predicted recoil using
Eq.~(\ref{eq:generic}), as well as our original empirical formula,
and compare the probabilities for observing a recoil in a given
velocity range. Our results are summarized in Fig.~\ref{fig:rec} and
Table~\ref{tab:rec}. The dramatic effect of the new recoil prediction
is apparent. Recoils in the range $1000\ \KMS - 3000 \KMS$ are
significantly enhanced, leading to a realistic chance for observing
large recoils. Details on how the statistical studies were performed
can be found in~\cite{Lousto:2009ka}.

\begin{table}
\caption {Probability that the recoil velocity will be in a given
range $P$, and the probability that the recoil will be a in 
a given range along the line of sight $P_{\rm obs}$ (to relate
to possible redshift measurements in galaxies) for the new
prediction (LEFT) and the old prediction (RIGHT)}
\label{tab:rec}
\begin{ruledtabular}
\begin{tabular}{lcc|cc}
range & $P$ & $P_{\rm obs}$  & $P$ old & $P_{\rm obs}$ old \\
\hline
0-500     & 79.027\%  &   92.641\%   &        94.888\%   &    98.914\% \\
500-1000  & 15.399\%  &    6.177\%   &         4.921\%   &     1.067\% \\
1000-2000 &  5.384\%  &    1.164\%   &         0.191\%   &     0.019\% \\
2000-3000 &  0.189\%  &    0.018\%   &         0   &      0 \\
3000-4000 &  0.001\%  &    0.0001\% &        0      &    0
\end{tabular}
\end{ruledtabular}
\end{table}

\mysection{Conclusions and Discussion}
We revisited the scenario for the generation of large
gravitational radiation recoils acquired by the remnant BH after the merger of
BHBs and found that configurations with spins partially aligned with
the orbital angular momentum produce larger recoils (up to 1200 km/s
more) than those with spins lying in the orbital plane (aka superkicks). 
The new configuration maximizes the total momentum radiated 
by optimizing over the competing requirements of
maximizing the total power radiated (which occurs for the hangup
configuration) and the skew in the power
distribution (which is maximized in the superkick configuration).
Our
results imply a nonlinear coupling among components of the spins that
can be expressed in the simple form given by
Eq.~(\ref{eq:generic}).  Based on these new terms in the empirical
formula for recoils, we recalculate the probabilities for large recoils to
occur in astrophysical scenarios of BHB encounters when accretion
effects are included. At small angles of the
spins with the orbital angular momentum, the new term magnifies
recoil velocities by up to a factor 2.8 with respect to the previous
formula,
and we find non-negligible probabilities of observing black holes recoiling
at several thousand km/s, as reported in Table~\ref{tab:rec}. 
Our results indicate that there is  a need for
additional theoretical searches for large recoils in other
regions of the parameter space, and lend support
for additional observational searches for  high-velocity
black holes.

\acknowledgments

It is a pleasure to thank M.Volonteri and M.Dotti for discussions on
the spin and mass distributions of BHB mergers and making these
results available prior to publication.  We thank M.Campanelli and
J.Krolik for
valuable discussions on the manuscript.  We gratefully acknowledge the NSF for
financial support from Grants No. PHY-0722315, No. PHY-0653303,
No. PHY-0714388, No. PHY-0722703, No. DMS-0820923, No. PHY-0929114,
No. PHY-0969855, No. PHY-0903782, No. AST-1028087; and NASA for
financial support from NASA Grants No. 07-ATFP07-0158. 
Computational resources were provided by the Ranger
and Lonestar clusters at TACC (Teragrid allocation TG-PHY060027N) and
by NewHorizons at RIT.

\bibliographystyle{apsrev}
\bibliography{../../../../Bibtex/references,local}

\end{document}